\documentclass[preprint,aps,amsmath,superscriptaddress,nofootinbib]{revtex4}
\usepackage{bm}
\usepackage{epsfig}

\newif\ifpdf
\ifx\pdfoutput\undefined
\pdffalse 
\else
\pdfoutput=1 
\pdftrue
\fi

\def\Dsl{\hbox{/\kern-.6000em D}} 

\def\dsl{\,\raise.15ex\hbox{/}\mkern-13.5mu D}

\def\psip#1{\psi_{\mathbf{#1}}}
\def\chip#1{\chi_{\mathbf{#1}}}

\def\ltap{\ \raise.3ex\hbox{$<$\kern-.75em\lower1ex\hbox{$\sim$}}\ }
\def\gtap{\ \raise.3ex\hbox{$>$\kern-.75em\lower1ex\hbox{$\sim$}}\ }
\def\OMIT#1{}

\def\vsl{v\!\!\!\slash}

\def\OMIT#1{}

\newcommand{\nn}{\nonumber}

\newcommand{\bea}{\begin{eqnarray}}
\newcommand{\eea}{\end{eqnarray}}

\begin{document}

\ifpdf
\DeclareGraphicsExtensions{.pdf, .jpg}
\newcommand{\picspace}{\vspace{-2.5in}}
\newcommand{\picspacehalf}{\vspace{-1.75in}}
\else
\DeclareGraphicsExtensions{.eps, .jpg,.ps}
\newcommand{\picspace}{\vspace{0in}}
\newcommand{\picspacehalf}{\vspace{0in}}
\fi



\title{Doubly Heavy Baryons, Heavy Quark-DiQuark Symmetry and NRQCD} 

\author{Sean Fleming\footnote{Electronic address: fleming@physics.arizona.edu}}
\affiliation{Department of Physics, University of Arizona,
      	Tucson, AZ 85721
	\vspace{0.2cm}}

\author{Thomas Mehen\footnote{Electronic address: mehen@phy.duke.edu}}
\affiliation{Department of Physics, Duke University, Durham,  NC 27708\vspace{0.2cm}}
\affiliation{Jefferson Laboratory, 12000 Jefferson Ave., Newport News, VA 23606\vspace{0.2cm}}

\date{\today\\ \vspace{1cm} }



\begin{abstract}

In the heavy quark limit, properties of heavy mesons and doubly heavy baryons are related by heavy quark-diquark symmetry. This
problem is reanalyzed in the framework of Non-Relativistic QCD (NRQCD). We introduce a novel method for deriving Potential NRQCD
(pNRQCD) Lagrangians for composite fields from vNRQCD, which contains quarks and antiquarks as explicit degrees of freedom and
maintains manifest power counting in the velocity via a label formalism. A Hubbard-Stratonovich transformation is used to
eliminate four quark interactions in vNRQCD and then quarks and antiquarks are integrated out to get effective Lagrangians for
composite fields. This method is used to rederive Lagrangians for the $Q\bar Q$ and $QQ$ sectors of pNRQCD and 
give a correct derivation of the $O(1/m_Q)$ prediction for the  hyperfine splitting of doubly heavy baryons.

\end{abstract}
\maketitle

\newpage

Mesons containing a single heavy quark 
$(Q{\bar q})$ and baryons containing two heavy antiquarks $({\bar Q} {\bar Q} {\bar q})$ are related
by quark-diquark symmetry. The two heavy antiquarks experience an attractive 
Coulomb force which can be computed in perturbation theory and the ground state 
of this two body system is a diquark in the $\bf 3$ of color with spin 1.
The mass of the diquark is very large and the spatial extent of the diquark is small, $r \sim 1/(m_Q v) \ll 
1/\Lambda_{\rm QCD}$, thus the light degrees of freedom in the baryon orbit 
a static source of color in the $\bf 3$ representation. The light degrees of freedom 
in the heavy meson also orbit a static source of color in the $\bf 	3$ representation,
therefore the configuration of the light degrees of freedom in the two hadrons is identical. 
The diquark and quark, however, have different spin-dependent interactions which are suppressed 
in the heavy quark limit. Furthermore there are corrections coming from the composite 
nature of the diquark. 

In Ref.~\cite{Savage:di} these ideas were implemented in the framework of 
heavy quark effective theory (HQET) by writing an effective Lagrangian for heavy quarks 
and diquarks which exhibits a $U(5)$ superflavor symmetry that permutes the five spin states 
of the heavy quark and diquark: 
\bea \label{hqlag}
{\cal L} =h^\dagger \, iD_0 \, h +{\bf T}^\dagger \, i D_0 \, {\bf T}   
+\frac{g_s}{2 m_Q} h^\dagger \, \bm{\sigma} \cdot {\bf B} \, h 
- \frac{i g_s}{2 m_Q} {\bf T}^\dagger \cdot {\bf B} \times {\bf T} 
+ ...  \, .
\eea
Here $h$ is the spin-$\frac{1}{2}$ heavy quark field, ${\bf T}$ the spin-1 diquark field, and  ${\bf B}$ the
chromomagnetic field. The first two terms are the leading order Lagrangian  which is manifestly $U(5)$ symmetric. The last two
terms are the heavy quark spin-symmetry breaking  $O(1/m_Q)$ corrections which are responsible for the hyperfine splittings of the
ground state heavy mesons and doubly heavy baryons. The ellipsis represents the kinetic energy terms which are $O(1/m_Q)$ but do
not explicitly break the heavy quark spin symmetry as well as higher order terms. The coefficient of the diquark chromomagnetic
coupling differs from Ref.~\cite{Savage:di} by a factor of 1/2 which will be explained below. The ground state heavy quark doublet
consists of a spin-0 meson, $P$, and spin-1 meson, $P^*$, and the ground state doubly heavy baryon doublet consists of  a spin-1/2
baryon, $\Sigma$, and spin-3/2 baryon, $\Sigma^*$. Matrix elements of the spin-symmetry breaking operators in Eq.~(\ref{hqlag})
can be computed using the methods of Ref.~\cite{Savage:di} to make the following prediction for the hyperfine splittings:
\bea \label{hf}
m_{\Sigma^*}-m_\Sigma = \frac{3}{4}(m_{P^*}-m_P) \, .
\eea 
This result is also easily derived in the quark model, see e.g. \cite{Lewis:2001iz}. It can also be generalized to baryons with
more than one heavy flavor, but since there is no experimental evidence for $bcq$ baryons to date, we will restrict our discussion
to the case of only one heavy quark flavor. 

Since the publication of Ref.~\cite{Savage:di} it has become clear that HQET is not the correct 
effective theory for hadrons containing two or more heavy quarks (or anti-quarks). This is
because HQET is formulated as an expansion in $\Lambda_{\rm QCD}/m_Q$, while a bound state containing
two or more heavy quarks has two additional scales, $m_Q v$, the typical momentum of the heavy 
quarks within the bound state, and $m_Q v^2$, the typical kinetic energy of the heavy quarks.
The appropriate theory for dealing with problems involving two heavy quarks is Non-Relativistic Quantum Chromodynamics 
(NRQCD)\cite{Bodwin:1994jh,Luke:1999kz}. This theory has the same Lagrangian as HQET but 
is formulated as an expansion in $v$ rather than $\Lambda_{\rm QCD}/m_Q$. The  goal 
of this paper is to re-derive the Lagrangian in Eq.~(\ref{hqlag}) and thereby the prediction in 
Eq.~(\ref{hf}) using NRQCD as a starting point. We will also discuss higher order
corrections to doubly heavy baryon hyperfine splittings.

When Ref.~\cite{Savage:di} was published the result in Eq.~(\ref{hf}) was of academic interest because no doubly heavy baryons had been
observed. Recently the SELEX experiment has published evidence for a number of $ccu$ and $ccd$
states~\cite{Mattson:2002vu,Moinester:2002uw,Ocherashvili:2004hi}.  One may worry about the applicability of our results to doubly charm baryons
since NRQCD assumes a hierarchy of scales  $m_Q \gg m_Q v \gg m_Q v^2$, while in bound states with two $c$ quarks or a $c$
and $\bar c$ quark these scales are not so widely separated: $m_c \sim 1.4$ GeV, $m_c v \sim 750$ MeV and $m_c v^2\sim \Lambda_{\rm QCD} \sim
400$ MeV. There are cases where the $v^2$ expansion  converges very poorly for charmonium, see for example the calculations of 
Refs.~\cite{Gremm:1996df,Braaten:2002fi}, and we suspect this will also happen for some observables involving doubly charmed baryons. In fact,
perturbative predictions for production rates \cite{Kiselev:2001fw}  and  weak decays \cite{Guberina:1999mx} seriously disagree with SELEX
data~\cite{Mattson:2002vu,Moinester:2002uw,Ocherashvili:2004hi}. However, using the known $D^*$ and $D$ masses in Eq.~(\ref{hf}) yields
$m_{\Sigma^*}-m_\Sigma \approx 105$ MeV which deviates from the experimental value of $78$ MeV by only  $\sim 30$\%,  which is the size of the
expected  corrections. Lattice calculations of doubly heavy  baryon hyperfine splittings are also approximately $80$
MeV~\cite{Lewis:2001iz}. This makes it seem plausible that the $v$ expansion could be effective for this observable.

The main point of the paper is to understand how approximate heavy quark-diquark symmetry arises within the framework of NRQCD and to lay the ground work for
computing systematic corrections to the leading order prediction in Eq.~(\ref{hf}) which are clearly necessary for understanding the  spectrum of doubly charm
baryons. An important aspect of this paper is our novel method for deriving the Potential NRQCD (pNRQCD) Lagrangian directly from the NRQCD Lagrangian of Luke,
Manohar and Rothstein~\cite{Luke:1999kz}, which is sometimes referred to as vNRQCD.  In the $Q\bar Q$ sector, auxiliary color-singlet and color-octet fields
are introduced via a Hubbard-Stratonovich transformation to eliminate four-quark interaction terms from the vNRQCD Lagrangian.  The quark and antiquark of
vNRQCD are integrated out of the theory, leaving an effective action for the color-singlet and color-octet fields and usoft gluons which corresponds to the
leading action of pNRQCD~\cite{Brambilla:1999xf}. This helps clarify the relationship between the two formulations of NRQCD at the level of the classical
Lagrangian.  

In section I, we rederive the pNRQCD action in the $Q\bar Q$ sector to $O(v)$ using our formalism. In section II, we extend this formalism to the $QQ$ sector
and derive the leading order spin symmetry breaking couplings of color antitriplet and sextet diquarks to usoft chromomagnetic fields. Section III contains a
discussion of soft gluons and how are these are handled in our formalism. We also discuss higher order corrections to Eq.~(\ref{hf}). Conclusions are given in
Section IV. While this work was in preparation, a pNRQCD treatment of baryons with two and three heavy quarks	 appeared
in Ref.~\cite{Brambilla:2005yk}. Their prediction for the hyperfine splitting agrees with Eq.~(\ref{hf}).
They also consider diquarks with two flavors of heavy quark and give the pNRQCD Lagrangian to $O(v^2)$. 
 
\section{pNRQCD from vNRQCD}

In this section, we describe our method for obtaining an effective action for composite fields  from the vNRQCD Lagrangian by introducing
auxiliary fields and integrating out heavy quarks and antiquarks. We illustrate our method by rederiving the pNRQCD Lagrangian
\cite{Brambilla:1999xf}, which is formulated in terms of composite color-singlet and color-octet fields.

Our starting point is the formulation of NRQCD due to Luke, Manohar, Rothstein~\cite{Luke:1999kz}. The effective theory contains several
different fields: heavy quarks, whose kinetic energy and momentum are $O(m_Q v^2)$ and $O(m_Q v)$, respectively, soft quarks and gluons
whose  energy and momentum is $O(m_Q v)$ and ultra-soft (or usoft) quarks and gluons whose  energy and momentum is $O(m_Q v^2)$. The scales
$m_Q$ and $m_Q v$ are removed from the effective theory using the trick of rephasing the NRQCD fields so that derivatives on the effective
field theory fields bring only $O(m_Q v^2)$ momentum~\cite{Luke:1999kz,Georgi:1990um}.  Thus, soft fields and heavy quarks will carry a
three vector label which indicates the  $O(m_Q v)$ momentum carried by these fields, while usoft fields carry no such label. We refer the reader
to Ref.~\cite{Luke:1999kz}  for a detailed discussion of notation, power counting, and the renormalization group in this theory. 

Any light degrees of freedom in a quarkonium or a doubly heavy baryon have energy and momentum $O(\Lambda_{\rm QCD})$. For the physically
irrelevant limit of extremely heavy quarks $\Lambda_{\rm QCD} \ll m_Q  v^2$, while for charmonium or doubly charmed baryons we have
$\Lambda_{\rm QCD} \sim m_Q v^2$. In either case, the light degrees of freedom are described by usoft fields.  Soft degrees of freedom appear
only in loops when computing matrix elements between heavy quarkonia or doubly heavy baryon states.  Therefore, it is possible to integrate out
the soft loops and derive an effective action for only usoft and heavy degrees of freedom. This is essentially the philosophy of (weakly
coupled) pNRQCD~\cite{Brambilla:1999xf}. 

The validity of effective field theory containing only usoft and heavy degrees of freedom is not so clear in the context of the velocity renormalization group
introduced in Ref.~\cite{Luke:1999kz}. In this method for summing logs of $v$, the renormalization scale of soft loops and usoft loops are correlated:  $\mu_s =
m_Q \nu^2$ and $\mu_{\rm us} = m_Q \nu$, and $\nu$ is evolved  from $\nu = 1$ to $\nu = v$. As argued in Ref.~\cite{Manohar:2000mx}, both soft loops and usoft
loops must be included throughout  the evolution to correctly resum logarithms of $v$.  Thus in vNRQCD, one must integrate out soft gluons at the scale $m_Q v^2$,
after summing logarithms of $v$.  However, in the pNRQCD approach to summing logarithms~\cite{Brambilla:1999xf}, soft gluons are integrated out at the
intermediate scale $m_Q v$.  We will not be concerned with resumming logs of $v$  and  so this issue should not affect the results of this paper.  

The leading order vNRQCD Lagrangian in the $Q\bar{Q}$ sector is 
\bea \label{lo}
{\cal L} &=& -\frac{1}{4}F_{\mu \nu} F^{\mu \nu} 
+ \sum_{\bf p} \psi_{\bf p}^\dagger \left( i D_0 -\frac{({\bf p} - i {\bf D})^2}{2 m_Q}\right) \psi_{\bf p}
+ \sum_{\bf p} \chi_{\bf p}^\dagger \left( i D_0 -\frac{({\bf p} - i {\bf D})^2}{2 m_Q}\right) \chi_{\bf p}
\nn \\
&-&  \sum_{\bf p,q} \frac{g_s^2}{({\bf p} - {\bf q})^2}\psi_{\bf q}^\dagger T^a \psi_{\bf p}
\chi_{\bf -q}^\dagger \bar{T}^a \chi_{\bf -p} + ... \, ,
\eea 
where the ellipsis represents higher order corrections and terms including soft gluons. We use a Fierz  transformation to project the potential
onto color-singlet and color-octet channels and obtain position space potentials by Fourier tranforming with respect to the labels:
\bea
\frac{g_s^2}{({\bf p - q})^2} = \int d^3 {\bf r} \frac{g_s^2}{4 \pi r}e^{i \bf (p-q)\cdot r} \, .
\eea
The Lagrangian is then
\bea
{\cal L} &=&  
 -\frac{1}{4}F^{\mu\nu}F_{\mu \nu} + \sum_{\bf p}
 \psip p ^\dagger   \left( i D^0 - \frac{\left({\bf p}-i{\bf D}\right)^2}{2 m_Q}  \right) 
 \psip p  + \sum_{\bf p} \chi_{\bf p}^\dagger \left( i D_0 -\frac{({\bf p} - i {\bf D})^2}{2 m_Q}\right) \chi_{\bf p}
\\
&&-  \int d^3  {\bf r} \, V^{(1)}(r) \left( \sum_{\bf q} e^{-i \bf q\cdot r} 
\frac{1}{\sqrt{N_c}}\, \psi^\dagger_{\bf q} \, \chi^\dagger_{\bf -q}\right)
\left( \sum_{\bf p} e^{i \bf p\cdot r} 
\frac{1}{\sqrt{N_c}} \, \chi_{\bf -p} \, \psi_{\bf p}\right) \nn \\
&&-  \int d^3  {\bf r} \, V^{(8)}(r) \left( \sum_{\bf q} e^{-i \bf q \cdot r} 
\sqrt{2} \,\psi^\dagger_{\bf q}\,  T^a \,\chi^\dagger_{\bf -q}\right)
\left( \sum_{\bf p} e^{i \bf p\cdot r} 
\sqrt{2} \,\chi_{\bf -p}\, T^a \,\psi_{\bf p}\right) \, ,\nn 
\eea 
where 
\bea
V^{(1)}(r) = - C_F\frac{\alpha_s}{r}, \qquad V^{(8)}(r) = \left(\frac{C_A}{2}-C_F\right) \frac{\alpha_s}{r} \, .
\eea
Next we introduce auxiliary fields into the Lagrangian via a Hubbard-Stratonovich transformation.
We add to the Lagrangian 
\bea 
\Delta {\cal L} &=& \int d^3  {\bf r}  \, V^{(1)}(r) \, \left(S_{\bf r}^\dagger - \sum_{\bf q} e^{-i \bf q\cdot r} 
\frac{1}{\sqrt{N_c}}\, \psi^\dagger_{\bf q} \, \chi^\dagger_{-{\bf q}}\right)  
\left(S_{\bf r} -  \sum_{\bf p} e^{i \bf p\cdot r} 
\frac{1}{\sqrt{N_c}} \, \chi_{\bf -p} \, \psi_{\bf p}\right) \nn \\
&& +\, V^{(8)}(r) \, \left( O^{a \dagger}_{\bf r} - \sum_{\bf q} e^{-i \bf q\cdot r} 
\sqrt{2} \,\psi^\dagger_{\bf q}\,  T^a \,\chi^\dagger_{-{\bf q}}\right)
\left( O^a_{\bf r} - \sum_{\bf p} e^{i \bf p\cdot r} 
\sqrt{2} \,\chi_{\bf -p}\, T^a \,\psi_{\bf p}\right) \, .
\eea
Note that the variable ${\bf r}$ is a continuous variable rather than a label. 
The terms quartic in the quark and antiquark fields cancel by construction and we are left with 
\bea\label{l1}
{\cal L} &=& -\frac{1}{4}F_{\mu \nu} F^{\mu \nu} 
+ \sum_{\bf p} \psi_{\bf p}^\dagger \left( i D_0 -\frac{({\bf p} - i {\bf D})^2}{2 m_Q}\right) \psi_{\bf p}
+ \sum_{\bf p} \chi_{\bf p}^\dagger \left( i D_0 -\frac{({\bf p} - i {\bf D})^2}{2 m_Q}\right) \chi_{\bf p}
 \\
&+& \int d^3 \, {\bf r} V^{(1)}({\bf r}) \left( S^\dagger_{\bf r} S_{\bf r} 
- S^\dagger_{\bf r}\sum_{\bf p} e^{i {\bf p \cdot r}} \frac{1}{\sqrt{N_c}} \chi_{- {\bf p}}\psi_{{\bf p}}
- S_{\bf r}\sum_{\bf q} e^{-i {\bf q \cdot  r}} \frac{1}{\sqrt{N_c}} \psi^\dagger_{{\bf q}}\chi^\dagger_{-{\bf q}}
\right) \nn \\
&+& \int d^3 \, {\bf r} V^{(8)}({\bf r}) \left( O^{a \dagger}_{\bf r} O^a_{\bf r} 
- O^{a \dagger}_{\bf r}\sum_{\bf p} e^{i {\bf p\cdot  r}} \sqrt{2} \chi_{- {\bf p}} T^a \psi_{{\bf p}}
- O^a_{\bf r}\sum_{\bf q} e^{-i {\bf q\cdot  r}} \sqrt{2} \psi^\dagger_{{\bf q}} T^a \chi^\dagger_{-{\bf q}}
\right) \, .\nn
\eea 
Integrating out the fields $S_{\bf r}$ and $O^a_{\bf r}$ yields the original NRQCD Lagrangian. Integrating out the fields $\psi_{\bf p}$ and
$\chi_{\bf p}$ will yield an effective action  for the fields $S_{\bf r}$ and $O^a_{\bf r}$ which can be used to study quarkonium at low
energies. The terms in the Lagrangian of Eq.~(\ref{l1}) which involve the quark and antiquark fields are reparametrization 
invariant~\cite{Luke:1992cs,Luke:1999kz}. The terms involving the auxiliary fields are not, but can be made reparametrization invariant 
by making the following substitution:
\bea \label{rpi}
e^{i {\bf p\cdot  r}}  \chi_{- {\bf p}} (T^a) \psi_{{\bf p}} \to 
e^{i {\bf p\cdot  r}} \left( \chi_{- {\bf p}} e^{-{\bf r\cdot  D}/2} \right) (T^a)
\left( e^{{\bf r\cdot  D}/2} \psi_{{\bf p}} \right) \, .
\eea
In this expression both covariant derivatives act only on the fields inside their respective parantheses and in the first set of parentheses 
the covariant derivative acts on $\chi_{-\bf p}$ from the right.  In NRQCD power counting, ${\bf p} \sim m_Q v$ so we must take  ${\bf r}
\sim (m_Q v)^{-1}$. As a result ${\bf r\cdot  D} \sim v$ and the factors $e^{\pm \bf {r\cdot  D}/2}$ in the Lagrangian can be expanded in powers of
$\bf r\cdot  D$. The $O(v)$ term in this expansion includes  a coupling to usoft gluon fields. When we integrate out the $\psi_{\bf p}$ and
$\chi_{\bf p}$ fields, this coupling  is needed to derive the couplings of the composite fields to usoft chromoelectric fields  at $O(v)$.

In passing we note that we can rewrite the exponentials of covariant derivatives using Wilson lines.
We define 
\bea \label{Wilson}
W_{\bf r}({\bf x}) = P \exp \left(i g_s \int_{- \infty}^0 ds \, {\bf r} \cdot {\bf A}({\bf x} + s {\bf r})\right) \, ,
\eea
which obviously obeys ${\bf r \cdot D} \, W_{\bf  r}(x) =0$. Then we have 
\bea 
{\bf r \cdot D} W_{\bf r}({\bf x}) f({\bf x}) =  W_{\bf  r}({\bf x}) {\bf r \cdot \partial} f({\bf x}) \, ,
\eea
which is equivalent to the operator relation  
\bea 
{\bf r\cdot D} =  W_{\bf r}({\bf x}) \, {\bf r\cdot \partial} \, W^\dagger_{\bf r}({\bf x}) \,,
\eea
where ${\bf r\cdot D}$ and ${\bf r\cdot \partial}$ act only to the right. This allows us to write
\bea 
\left( e^{{\bf r \cdot D}/2} \psi_{{\bf p}}({\bf x}) \right) &=& W_{\bf r}({\bf x}) 
\left( e^{{\bf r \cdot \partial}/2} W^\dagger_{\bf r}({\bf x})\psi_{{\bf p}}({\bf x}) \right)\nn \\
&=& W_{\bf r}({\bf x})  W^\dagger_{\bf  r}({\bf x} +{\bf r}/2)\psi_{{\bf p}}({\bf x}+{\bf r}/2) \nn \\
&=& W_{\bf r}({\bf x},{\bf x} +{\bf r}/2)\psi_{{\bf p}}({\bf x}+{\bf r}/2) \, ,
\eea
where 
\bea
 W_{\bf r}({\bf x},{\bf x} +{\bf r}/2) = 
 P \exp \left(i g_s \int_{0}^1 ds \, \frac{{\bf r}}{2} \cdot {\bf A}\left({\bf x} +\frac{{\bf r}}{2} -s \frac{{\bf r}}{2}\right)\right) \, ,
\eea
is a Wilson line connecting the points $\bf x$ and $\bf x +r/2$. Then the equations of motion for $S_{\bf r}({\bf x})$ and 
$O^{a}({\bf x})$ yield
\bea
S_{\bf r}({\bf x}) &=& \sum_{{\bf p}} \frac{1}{\sqrt{N_c}}e^{i \bf p \cdot r} \chi_{-\bf p}({\bf x- r}/2)\,
W_{\bf r}({\bf x- r}/2,{\bf x + r}/2) \, \psi_{\bf p}({\bf x+ r}/2) \nn \\
O^a_{\bf r}({\bf x}) &=& \sum_{{\bf p}} e^{i \bf p \cdot r} \chi_{-\bf p}({\bf x- r}/2) \,
W_{\bf r}({\bf x- r}/2,{\bf x}) \,\sqrt{2} T^a \, W_{\bf r}({\bf x},{\bf x + r}/2) \,\psi_{\bf p}({\bf x+ r}/2) \, . \nn
\eea
We see that once the corrections which restore reparametrization invariance are included to all orders the fields $S_{\bf r}({\bf x})$ are
nonlocal products of quark-antiquark fields separated by a distance $\bf r$ with center of mass coordinate $\bf x$. The Wilson lines ensure that
the nonlocal fields transform correctly under usoft gauge transformations, namely as 
\bea 
\psi_{\bf p} \to V \psi_{\bf p} \qquad S_{\bf r} \to S_{\bf r} \qquad O^a_{\bf r} \to {\cal V}^{ab} \,O^b_{\bf r} \, ,
\eea
where $V$ is in the fundamental representation and ${\cal V}^{ab}$ is in the adjoint with $V^\dagger T^a V = {\cal V}^{ab}
T^b$. The composite fields are very similar to interpolating fields introduced in Ref.~\cite{Brambilla:1999xf} for the 
construction of pNRQCD. Because the separation $\bf r$ is a short distance scale compared to the usoft scale
we can perform a multipole expansion in $\bf r$.

\begin{figure}[!t]
  \centerline{\epsfysize=6.0truecm \epsfbox[100 430 560 685]{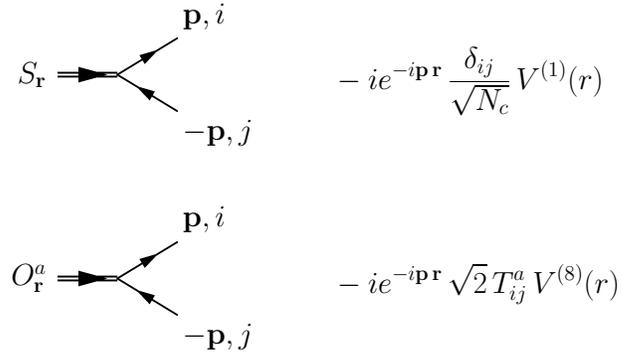}  }
 {\tighten
\caption[1]{Feynman rules for the coupling of $S_{\bf r}$ and $O^a_{\bf r}$ to quarks and antiquarks}
\label{fr0} }
\end{figure}

We will first derive the kinetic terms for the $S_{\bf r}$ and $O^a_{\bf r}$ fields, for which we can neglect usoft gauge fields.
The Feynman rules for the couplings of these composite fields to quarks and antiquarks  are shown in Fig.~\ref{fr0}. In these
diagrams ${\bf p}$ and $-{\bf p}$ are the label momentum of the quark and antiquark, respectively. The vertex is nonvanishing only
when the labels of the quark and antiquark fields are equal and opposite.  An effective action for
the $S_{\bf r}$ and $O^a_{\bf r}$ fields is obtained by integrating out the fields $\psi_{\bf p}$ and  $\chi_{\bf p}$, which amounts to
doing  one-loop graphs with quark anti-quark fields on the internal lines and an arbitrary number of external auxiliary field
lines. This is similar to the theory of the superconductivity in which one introduces a pairing field and then integrates out the
fermions to get an effective potential for the gap, see Ref.~\cite{Weinberg:1996kr}. However, in this case we are only interested
in processes involving a single $Q \bar Q$ pair and therefore only need the part of the action that is quadratic in the composite
fields. Therefore, computation of the effective action requires a one loop diagram with two external auxiliary fields, $S^\dagger
S$ or $O^{a \dagger} O^a$. (We will find terms with $S^\dagger O^a$ or $O^{a \dagger} S$ once we include external usoft gluons.)

\begin{figure}[!h]
  \centerline{\epsfysize=4.0truecm \epsfbox[120 560 500 680]{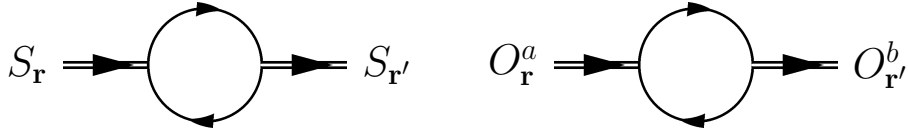}  }
 {\tighten
\caption[1]{One loop contributions to effective action for $S_{\bf r}$ and $O^a_{\bf r}$ fields.}
\label{ol}}
\end{figure}
The one loop graph in Fig.~\ref{ol} with two external composite fields, $S_{\bf r}$ and 
$S_{\bf r^\prime}$, yields 
\bea 
i m_Q \int \frac{d^3 {\bf l}}{(2\pi)^3} V^{(1)}(r) \frac{e^{i {\bf l \cdot (r-r^\prime)}}}{{\bf l}^2-m_Q E - i \epsilon} 
V^{(1)}(r^\prime) \, .
\eea 
In evaluating this graph the sum over virtual labels and integral over virtual
residual momentum have been combined into a single integral.
The one loop diagram with external composite fields $O^a_{\bf r}$ and $O^b_{\bf r^\prime}$  shown in Fig.~\ref{ol}
yields a similar result. The effective Lagrangian obtained from integrating out the quark and antiquark fields is given by 
\bea \label{eff1}
{\cal L} &=&
\int d^3{\bf r} d^3 {\bf r^\prime} S_{\bf r}^\dagger \left( V^{(1)}(r) \delta^3({\bf r-r^\prime}) +
\int \frac{d^3 {\bf l}}{(2\pi)^3} V^{(1)}(r) \frac{e^{i {\bf l \cdot (r-r^\prime)}}}{{\bf l}^2/m_Q- E - i \epsilon} V^{(1)}(r^\prime) \right) 
S_{\bf r^\prime} \\
&+& \int d^3{\bf r} d^3{\bf r}^\prime O_{\bf r}^{a\dagger} \left( V^{(8)}(r) \delta^3({\bf r-r^\prime}) +
\int \frac{d^3 {\bf l}}{(2\pi)^3} V^{(8)}(r) \frac{e^{i {\bf l \cdot (r-r^\prime)}}}{{\bf l}^2/m_Q- E - i \epsilon} V^{(8)}(r^\prime) \right) 
O^a_{\bf r^\prime} \nn \, .
\eea 
This Lagrangian is most easily interpreted using the following notation. We think of the fields $S_{\bf r}, O^a_{\bf r}$ as vectors on a Hilbert space spanned
by eigenkets of $\bf r$, so we write $S_{\bf r} = \langle {\bf r} |S\rangle$. The fields also depend on $x$ but this dependence will be suppressed for
simplicity. We define operators $\hat V^{(1,8)}$ whose matrix elements are $\langle {\bf r}^\prime |\hat V^{(1,8)}|{\bf r}\rangle = V^{(1,8)}(r)\delta^3({\bf
r-r^\prime})$, momentum eigenkets, defined by $\langle {\bf r} | {\bf l}\rangle = e^{i {\bf l \cdot r}}$ and a free Hamiltonian, $H_0 |{\bf l}\rangle =
{\bf l}^2/m_Q|{\bf l}\rangle$. In this quantum mechanical notation, the action can be written as
\bea \label{eff2}
{\cal  L} &=&
\langle S|{\hat V}^{(1)} - {\hat V}^{(1)} \frac{1}{\hat{E}-H_0}{\hat V}^{(1)}|S\rangle +
\langle O^a|{\hat V}^{(8)} - {\hat V}^{(8)} \frac{1}{\hat{E}-H_0}{\hat V}^{(8)}|O^a\rangle \,,
\eea
where $\hat{E} = i \partial /\partial t$.
Inserting complete sets of states $1 = \int d^3r | {\bf r} \rangle \langle {\bf r} |=
\int \frac{d^3l}{(2\pi)^3}| {\bf l} \rangle \langle {\bf l} |$ in the appropriate places 
in Eq.~(\ref{eff2}) allows one to reproduce Eq.~(\ref{eff1}). Using algebra, it is easy
to show 
\bea \label{simple}
\langle \phi |V - V\frac{1}{\hat{E}-H_0} V| \phi \rangle &=&  \langle \phi |\hat{E}-H - (\hat{E}-H)\frac{1}{\hat{E}-H_0}(\hat{E}-H)| \phi \rangle \\
&=& \langle \phi |V\frac{1}{\hat{E}-H_0}(\hat{E}-H)| \phi \rangle \nn \, .
\eea
where  $\phi$ denotes either composite field, $V$ is the relevant potential, and $H = H_0 + V$. The equation of motion is 
clearly $(\hat{E}-H)|\phi\rangle = 0$ or 
\bea
\left(i\partial_0 + \frac{\bm{\nabla}_{\bf r}^2}{m_Q} - V(r) \right) \phi(r) = 0 \,.
\eea
The second term on the right-hand side  of the first line in Eq.~(\ref{simple}) clearly vanishes by virtue of the equation of motion and can be removed
by a field redefinition. Then the action is given by 
\bea \label{eff3}
{\cal  L} &=&
\langle S|\hat{E}- H_0 - \hat{V}^{(1)}|S\rangle +
\langle O^a|\hat{E}- H_0 - \hat{V}^{(8)}|O^a\rangle \nn \\
&=&\int d^3 {\bf r} \, S_{\bf r}^\dagger \left( i \partial_0 + \frac{\bm{\nabla}_{\bf r}^2}{m_Q} -V^{(1)}(r)\right) S_{\bf r}
+ \int d^3 {\bf r} \, O_{\bf r}^{a \dagger} \left( i \partial_0 + \frac{\bm{\nabla}_{\bf r}^2}{m_Q} -V^{(8)}(r)\right) O^a_{\bf r} \, .
\eea
This is just the leading order Lagrangian of pNRQCD~\cite{Brambilla:1999xf}.

Next we consider corrections to the effective action that involve one loop diagrams with external
usoft gluons. One complication is that the couplings of the $A_0$ gluon to the elementary 
quark and antiquark fields that come from the covariant derivatives in the kinetic terms of 
Eq.~(\ref{lo}) are not suppressed by any powers of $v$. (Couplings to $A_i$ gluons are $v$ suppressed.) 
Naively, this suggests that one loop graphs with an arbitrary number of external
$A_0$ gluons must be summed to lowest order in $v$. Of course, one can go to $A_0=0$ gauge and all these
diagrams disappear. Therefore, the only corrections to the effective action that involve the $A_0$
gluons to lowest order in $v$ are simply the terms required to make the derivative acting on $O^a_{\bf r}$
in Eq.~(\ref{eff3}) covariant. In an arbitrary gauge there are cancellations between diagrams 
involving arbitrary gluons so that in the end the correction to the effective action is simple.
The calculation of the effective action is greatly simplified if a field redefinition is performed 
that removes the $O(v^0)$ interaction of the quark fields and the $A_0$ gluon. This of course
introduces new couplings to $A_0$ gluons in higher order $v$ suppressed interactions, but these
can be treated perturbatively.

Consider the kinetic terms for the quarks:
\bea
{\cal L} = 
\sum_{\bf p} \psi_{\bf p}^\dagger \left( i D_0 -\frac{({\bf p} - i {\bf D})^2}{2 m_Q}\right) 
\psi_{\bf p} \, .
\eea
To remove the $A_0$ coupling we we make the field redefinition~\cite{Bauer:2001yt}
\bea
\psi_{\bf p} = S_0 \tilde\psi_{\bf p}  \qquad S_0 = P \exp \left(-i g_s \int_{-\infty}^0 ds A_0(x +s)\right) 
\, ,
\eea
where $S_0$ is a Wilson line obeying $i D_0 S_0 = 0$. Therefore we have $S_0^\dagger i D_0 S_0 = i \partial_0$ and the kinetic term for the quarks in terms of
the new fields is:
\bea
{\cal L} = 
\sum_{\bf p} \tilde \psi_{\bf p}^\dagger \left( i \partial_0 -\frac{{\bf p}^2 -2 i{\bf p} S_0^\dagger i {\bf D} S_0
- S_0^\dagger {\bf D}^2 S_0}{2 m_Q}\right) \tilde \psi_{\bf p} \, .
\eea
We can write the terms with the Wilson lines more neatly if we use the following identities.
\bea 
S_0^\dagger {\bf D}_i S_0  = {\bf \nabla}_i + [S_0^\dagger {\bf D}_i S_0]
\equiv  {\bf \nabla}_i - i g {\bf B}_i  \equiv {\bf \tilde D}_i  \, .
\eea 
Brackets indicate that the covariant derivative acts only on terms inside the brackets.
The field ${\bf B}_i = \frac{i}{g} [S_0^\dagger {\bf D}_i S_0]$ is a gauge invariant generalization of the 
field ${\bf A}_i$. Written in terms of the new fields the action is 
\bea 
{\cal L} = 
\sum_{\bf p} \tilde \psi_{\bf p}^\dagger \left( i \partial_0 -\frac{({\bf p} - i {\bf \tilde D})^2}{2 m_Q}\right) \tilde \psi_{\bf p} \, ,
\eea
i.e., in terms of the new fields the Lagrangian is the same as before with the replacements
$\psi \to \tilde \psi, A_0 \to 0,{\bf A}_i \to {\bf B}_i$. When written in terms of the new fields the
the action has no $O(v^0)$ interactions with gluons. The terms involving auxiliary fields 
are 
\bea\label{aux}
&&\int d^3 \, {\bf r} V^{(1)}({\bf r}) \left( S^\dagger_{\bf r} S_{\bf r} 
- S^\dagger_{\bf r}\sum_{\bf p} e^{i {\bf p \cdot r}} \frac{1}{\sqrt{N_c}} \tilde \chi_{- {\bf p}} \tilde 
\psi_{{\bf p}}
- S_{\bf r}\sum_{\bf q} e^{-i {\bf q \cdot  r}} \frac{1}{\sqrt{N_c}}  \tilde\psi^\dagger_{{\bf q}} \tilde\chi^\dagger_{-{\bf q}}\right)  \\
&+& \int d^3 \, {\bf r} V^{(8)}({\bf r}) \left( O^{a \dagger}_{\bf r} O^a_{\bf r} 
- O^{a \dagger}_{\bf r}\sum_{\bf p} e^{i {\bf p\cdot  r}} \sqrt{2}  \tilde\chi_{- {\bf p}} {\cal Y}_0^{ab} T^b  \tilde\psi_{{\bf p}}
- O^a_{\bf r}\sum_{\bf q} e^{-i {\bf q\cdot  r}} \sqrt{2} \tilde\psi^\dagger_{{\bf q}} {\cal Y}_0^{ba} T^b \tilde\chi^\dagger_{-{\bf q}}
\right)  \nn \, .
\eea
The terms involving $S$ are the same as before once we make the substitutions $\psi \to \tilde \psi$ and
$\chi \to \tilde \chi$. The terms with $O^a$ contain an additional octet Wilson line, defined
by $S_0^\dagger T^a S_0 = {\cal Y}_0^{ab} T^b$. This octet Wilson line can be absorbed by redefinition 
of the composite field: $\tilde O^b = {\cal Y}_0^{ba} O^a$. It is straightforward to show that after the
field redefinition covariant derivatives ${\bf D}$ in Eq.~(\ref{rpi}) are replaced with $\tilde {\bf  D}$.
So the field redefinition can be implemented by making the following substitutions in the Lagrangian: 
\bea
\psi \to \tilde \psi, \, \chi \to \tilde \chi, \,  A_0 \to 0, \, {\bf A}_i \to {\bf B}_i, \,  S\to S 
, \,O^a \to \tilde O^a \, . 
\eea
When rewritten in terms of the new fields, the action contains no $v^0$ interactions involving gluons. We can now compute the
effective action to $O(v)$ for the fields $S, \tilde O^a$ and ${\bf B}_i$. Once this is completed the action can  be rewritten in
terms of the original fields by undoing the field redefinitions if desired. The result is equivalent to what is obtained if one
uses $A_0=0$ gauge and then imposes gauge invariance. Performing the field redefinitions makes it possible to maintain gauge
invariance at all stages of the calculation.

\begin{figure}[!ht]
  \centerline{\epsfysize=6.0truecm \epsfbox[100 475 530 730]{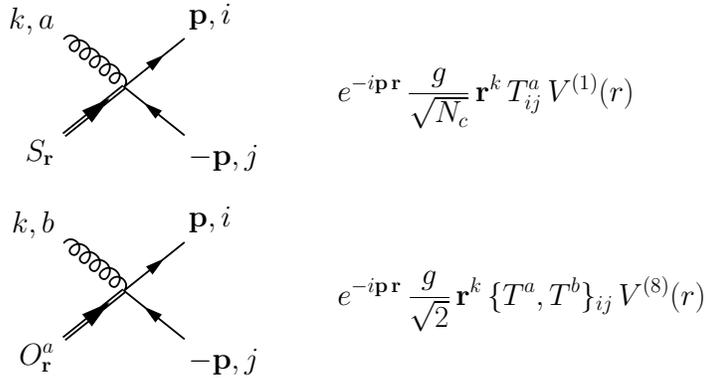}  }
 {\tighten
\caption[1]{$O(v)$ couplings of composite fields to external gluons, quark and anti-quark fields.}
\label{frv} }
\end{figure}
The relevant Feynman rules for graphs with one external usoft gluon, a quark, an anti-quark,
and a composite $S_{\bf r}$ or $O^a_{\bf r}$ field are shown in Fig.~\ref{frv}. 
These are obtained by extending the Lagrangian in Eq.~(\ref{l1}) to include the correction in Eq.~(\ref{rpi})
which restores reparametrization invariance and expanding to lowest order in the gluon fields.
There are also  $O(v)$ interactions in the coupling of quarks and antiquarks to usoft gluon fields.
There are no other $O(v)$ interactions in the Lagrangian at this order. At higher orders in $v$, it 
is no longer sufficient to consider only the reparametrization invariant extension of the Coulomb interaction in Eq.~(\ref{lo}).
At order $O(v^2)$ there are new potentials obtained from matching QCD onto pNRQCD~\cite{Manohar:2000hj}.

\begin{figure}[!ht]
  \centerline{\epsfysize=6.0truecm \epsfbox[110 450 500 670]{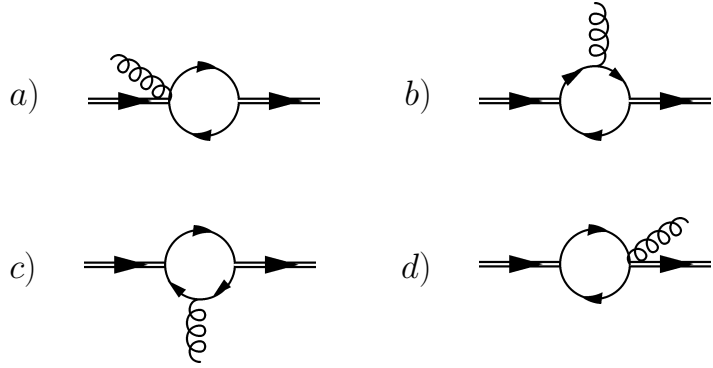}  }
 {\tighten
\caption[1]{$O(v)$ loop diagrams contributing to the coupling of composite fields and 
usoft gluon fields.}
\label{efv} }
\end{figure}
The four diagrams contributing to the $O(v)$  effective action for composite fields
are shown in Fig.~\ref{efv}. Because of color conservation at least one of the 
external fields must be an octet. We will describe the evaluation of the graph with 
external $S_{\bf r}$ and $O^a_{\bf r^\prime}$ in some detail. Let $E$
and $E^\prime$ be the energy of the $S_{\bf r}$ and $O^a_{\bf r^\prime}$ fields, respectively. 
Let the polarization index of the gluon be $i$ and its color index $b$. Working in the rest frame of $S_{\bf r}$
we find that graph b), in which the gluon connects directly to a quark line via the $\bf p \cdot A$ coupling 
in Eq.~(\ref{lo}), yields
\bea 
b) &=& i \frac{g}{m_Q}\frac{\delta^{ab}}{\sqrt{2N_c}} \int\frac{d^3{\bf l}}{(2\pi)^3}
\frac{V^{(1)}(r)}{E-{\bf l}^2/m_Q}\frac{V^{(8)}(r^\prime)}{E^\prime-{\bf l}^2/m_Q}
{\bf l}^i e^{-i{\bf l}\cdot {\bf r}} e^{i{\bf l}\cdot {\bf r^\prime}} \\
&=& i \frac{g}{m_Q}\frac{\delta^{ab}}{\sqrt{2N_c}}\int\frac{d^3{\bf l}}{(2\pi)^3}
\frac{V^{(1)}(r)}{E+\bm{\nabla}_{\bf r}^2/m_Q}\frac{V^{(8)}(r^\prime)}{E^\prime +\bm{\nabla}_{\bf r^\prime}^2/m_Q}
(- i \nabla^i_{\bf r^\prime}) e^{-i{\bf l}\cdot{\bf r}}
e^{i{\bf l}\cdot{\bf r^\prime}} \nn \\
&=& \frac{g}{m_Q}\frac{\delta^{ab}}{\sqrt{2N_c}}\nabla^i \, \delta^3({\bf r- r^\prime}) \nn \,.
\eea
The first line is obtained using the previously derived Feynman rules and performing the integral  over the energy in the loop by
contour integration. In the second line we have replaced factors  of the momentum appearing in the integrand with derivatives of
${\bf r}$ and ${\bf r^\prime}$ acting on the  exponential factors in the integrand. In the last line we used the equations of
motion to set $E+\bm{\nabla}_{\bf r}^2/m_Q = V^{(1)}(r)$ and $E^\prime +\bm{\nabla}_{\bf r^\prime}^2/m_Q = V^{(8)}(r^\prime)$, then
performed the integral over the loop momentum. Graph $c)$, in which the gluon is attached to the antiquark line, gives the same
result  as graph $b)$. Graph a) yields 
\bea
a) &=& \frac{g}{\sqrt{2N_c}}\delta^{ab} {\bf r}^i \int\frac{d^3{\bf l}}{(2\pi)^3}
\frac{V^{(1)}(r) \, V^{(8)}(r^\prime) \,e^{-i{\bf l}\cdot{\bf r}} \, e^{i{\bf l}\cdot{\bf r^\prime}} }{E^\prime -{\bf l}^2/m_Q} \\
&=& \frac{g}{\sqrt{2N_c}}\delta^{ab} {\bf r}^i \int\frac{d^3{\bf l}}{(2\pi)^3}
\frac{V^{(1)}(r) \, V^{(8)}(r^\prime)}{E^\prime +\bm{\nabla}_{\bf r^\prime}^2/m_Q}\,e^{-i{\bf l}\cdot{\bf r}} \,e^{i{\bf l} \cdot{\bf r^\prime} } \nn \\
&=& \frac{g}{\sqrt{2N_c}}\delta^{ab}\, {\bf r}^i \,V^{(1)}(r) \,\delta^3({\bf r- r^\prime}) \, .\nn
\eea
where we have used the Feynman rules of Fig.~\ref{frv} and the same manipulations used in the evaluation of graph
b). Graph d) yields 
\bea
-\frac{g}{\sqrt{2N_c}}\delta^{ab}\, {\bf r}^{\prime i} \,V^{(8)}(r^\prime)\, \delta^3({\bf r- r^\prime}) \, .
\eea
Combining all four diagrams then yields the following effective Lagrangian for the composite fields:
\bea
{\cal L}_v &=& -i \frac{g}{\sqrt{2N_c}}\int d^3{\bf r} \, d^3{\bf r}^\prime \tilde O^{\dagger a}_{{\bf r^\prime}}
\left[\frac{ 2 \bm{\nabla}_{{\bf r}^\prime }}{m_Q} +  
\, ({{\bf r}} \, V^{(1)}(r)- {{\bf r}^\prime} \,V^{(8)}(r^\prime))\right] \delta^3({\bf r}-{\bf r^\prime}) 
\cdot {{\bf B}}^a S_{\bf r} \\
&=& -i \frac{g}{\sqrt{2N_c}}\int d^3 {\bf r}  \left[ -\frac{2}{m_Q} 
[\bm{\nabla}_{{\bf r}} \tilde O^{\dagger a}_{\bf r}] S_{\bf r}+
{{\bf r}} \, (V^{(1)}(r)- V^{(8)}(r)) \tilde O^{\dagger a}_{\bf r} S_{\bf r}\right] 
\cdot{{\bf B}}^a  \, . \nn
\eea
In the second line we have integrated by parts and performed the integral over the $\delta$-function.
Now using the leading order equations of motion
\bea
V^{(1)}(r) S_{\bf r} &=& \left(i\partial_0 + \frac{\bm{\nabla}_{\bf r}^2}{m_Q}\right) S_{\bf r} \\
\tilde O^{\dagger a}_{\bf r} V^{(8)}(r) &=& \left(-i\partial_0 + \frac{\bm{\nabla}_{\bf r}^2}{m_Q}\right) 
\tilde O^{\dagger a}_{\bf r} \, ,
\eea 
and repeatedly integrating by parts one obtains 
\bea\label{dipole}
{\cal L}_v = -\frac{g}{\sqrt{2N_c}} \int d^3{\bf r} \, {{\bf r}} \cdot \, \partial_0 {{\bf B}}^a \, \tilde O^{\dagger a}_{\bf r} S_{\bf r} \, .
\eea
To write this in terms of the original fields, we first note
\bea 
\partial_0 {{\bf B}} = \frac{i}{g} \partial_0 [S_0^\dagger {{\bf D}} S_0]
= \frac{i}{g} S_0^\dagger [D_0,{{\bf D}}] S_0 = -S_0^\dagger {{\bf E}} S_0 \, ,\nn
\eea
where ${{\bf B}}$ and ${{\bf E}}$ are Lie algebra valued. In terms of components
this equation can be written as $\partial_0 {{\bf B}}^a = - {\cal Y}_0^{ba}\, {{\bf E}}^b$,
where ${\cal Y}_0^{ba}$ is the Wilson line defined shortly after Eq.~(\ref{aux}). We can use this to 
rewrite  Eq.~(\ref{dipole}) in terms of the original fields using  
${\cal Y}_0^{ba} \tilde O^{\dagger a}(r)= O^{\dagger a}(r)$.

Graphs of Fig.~\ref{efv} with two external octet fields, $O^a$ and $O^b$ and an external gluon with color
index $c$ are the same as graphs with one external singlet field except for the replacement of the color factor
$\delta^{ab}/\sqrt{2N_c} \to d^{abc}/2$. The $O(v)$  effective Lagrangian for the composite fields is
\bea \label{ov}
{\cal L}_v = g \sqrt{\frac{2}{N_c}} \int d^3 {\bf r} \,  
{\rm Tr}[O^{\dagger }_{\bf r} \, {\bf r} \cdot {\bf E} \,] \, S_{\bf r} \, + {\rm h.c.} 
+ g \int d^3{\bf r} \, {\rm Tr}[O^{\dagger }_{\bf r}\, \{ {{\bf r}} \cdot {{\bf E}}, O_{\bf r} \} ] \, .
\eea
Here we have defined $O_{\bf r} = O^a_{\bf r} T^a$ and ${{\bf E}} =  {{\bf E}}^a T^a$. This result is in agreement
with Ref.~\cite{Brambilla:1999xf}.

\section{An Effective Action for Composite Diquark Fields}

\begin{figure}[!t]
  \centerline{\epsfysize=6.0truecm \epsfbox[90 430 525 675]{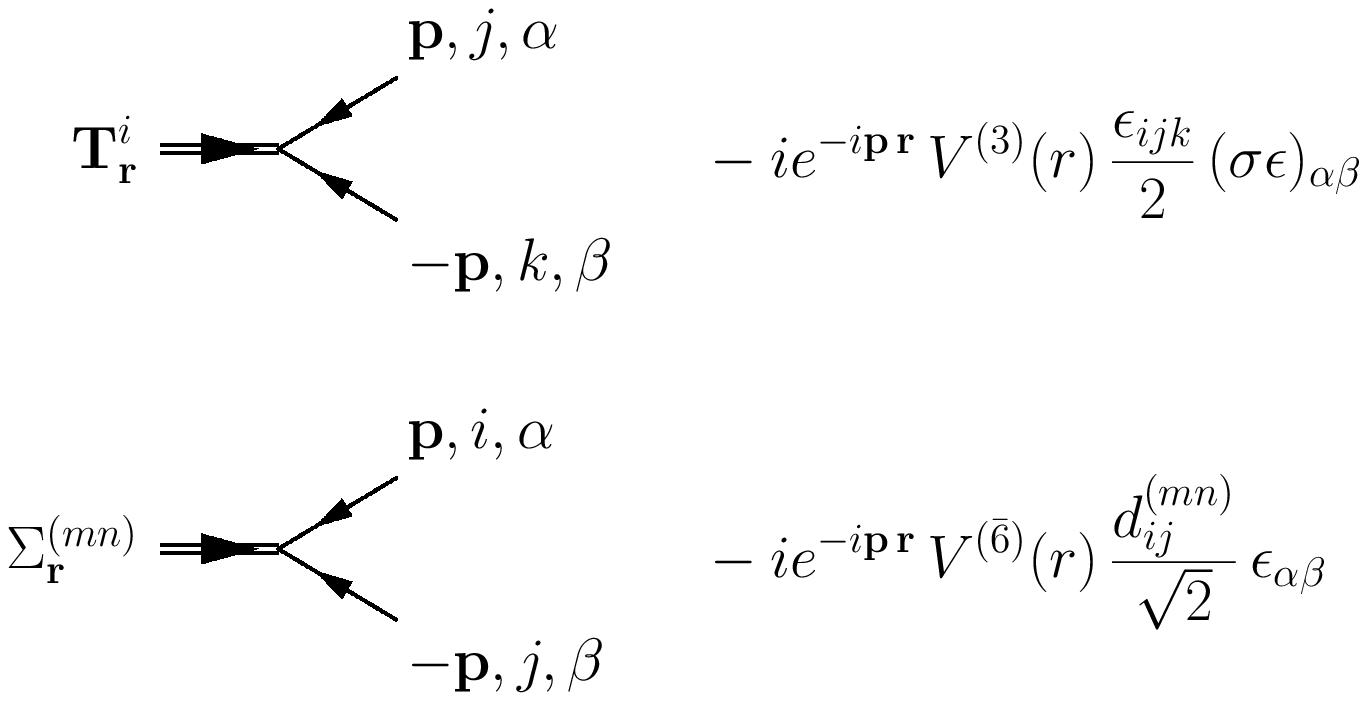}  }
 {\tighten
\caption[1]{Feynman rules for the coupling of the diquarks $ {\bf T}^i_{\bf r}$ and $\Sigma^{(mn)}_{\bf r}$ to antiquarks.}
\label{Fig5} }
\end{figure}

\begin{figure}[!t]
  \centerline{\epsfysize=6.0truecm \epsfbox[30 420 596 690]{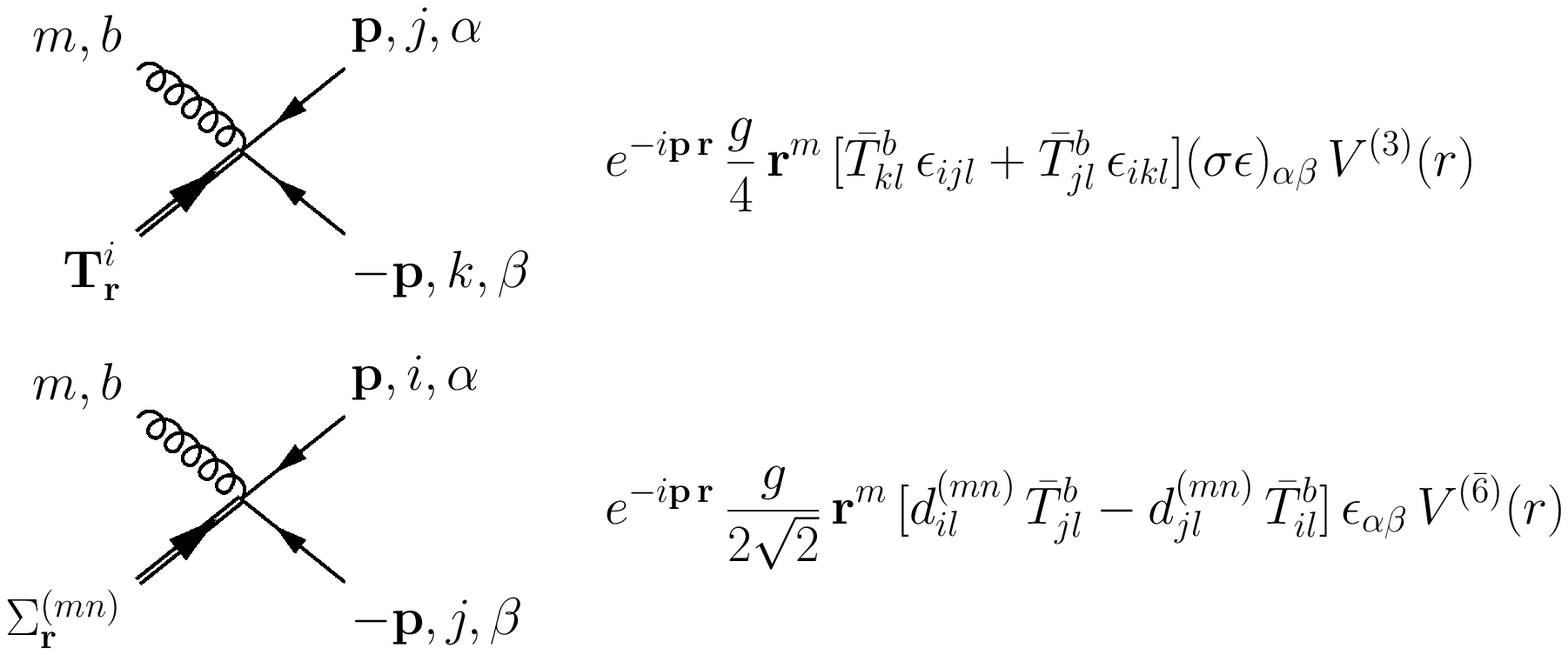}  }
 {\tighten
\caption[1]{$O(v)$ couplings of diquarks to gluons and quarks.}
\label{Fig6} }
\end{figure}

The NRQCD Lagrangian relevant to constructing an effective action for diquarks is 
\begin{eqnarray}\label{nrqcd}
{\cal L} &=&  
 -\frac{1}{4}F^{\mu\nu}F_{\mu \nu} + 
\sum_{\bf p} \chip p ^\dagger   \Biggl ( i D^0 - \frac{\left({\bf p}-i{\bf D}\right)^2}{2 m_Q} 
 + \frac{g}{2 m_Q} \,\bm{\sigma}\cdot {{\bf B}} 
 \Biggr )
 \chip p  \nonumber \\
&& - \frac{1}{2} \sum_{\bf p,q}
 \frac{g_s^2}{ \bf (p-q)^2} 
  \chip q ^\dagger \bar T^A \chip p \chip {-q}^\dagger \bar T^A \chip {-p} + \ldots \, .
\end{eqnarray}
The ellipsis represents higher order corrections as well as terms
including soft gluons. Also included is the leading spin symmetry violating interaction 
of the heavy antiquark since it is necesssary for generating hyperfine splittings. The 
following spin and color Fierz identities 
\bea\label{Fierz}
\delta_{\alpha \delta} \delta_{\beta \gamma} &=& -\frac{1}{2}(\sigma^i \epsilon)_{\alpha \beta} 
(\epsilon \sigma^i)_{\gamma \delta}  + \frac{1}{2} \epsilon_{\alpha \beta} \epsilon_{\delta \gamma} \nn \\
\bar{T}^a_{il}\bar{T}^a_{jk} &=& -\frac{2}{3}\sum_m \frac{1}{2}\epsilon_{mij} \epsilon_{mlk}
+\frac{1}{3}\sum_{(mn)} d^{\,(mn)}_{ij} d^{\,(mn)}_{kl}  \, ,
\eea 
are used to the project the potential onto the 
$\bf 3$ and $\bf \bar 6$ channels and to decompose the quark bilinears such as $\chip p \, \chip {-p}$
into operators with definite spin.
In the first line,  the Greek letters denote spin indices, the $\sigma^i$ denotes the Pauli matrices
and $\epsilon = i \sigma^2$. In the second line, Roman letters refer to the color indices and the 
matrices $d^{(mn)}_{ij}$ are symmetric  matrices in color space defined by
\bea
d^{(mn)}_{ij} = \left\{ 
\begin{array}{cc}
(\delta^m_i \delta^n_j + \delta^n_i \delta^m_j)/\sqrt{2} & m\neq n \\
\delta^m_i \delta^n_j & m = n
\end{array} \right. \, .
\eea
In denoting the matrices by $(mn)$ we do not distinguish $(mn)$ from $(nm)$ so there are six distinct matrices.
After Fourier tranforming with respect to the labels the Lagrangian can be rewritten as 
\bea\label{s1}
{\cal L} &=&  
 -\frac{1}{4}F^{\mu\nu}F_{\mu \nu} + 
 \sum_{\bf p} \chip p ^\dagger   \Biggl ( i D^0 - \frac{\left({\bf p}-i{\bf D}\right)^2}{2 m_Q} 
 + \frac{g}{2 m_Q} \, \bm{\sigma}\cdot {{\bf B}} 
 \Biggr )
 \chip p  \\
&-&\frac{1}{2} \int d^3 {\bf r} \, V^{(3)}(r) \left( \sum_{\bf q} e^{-i \bf q \cdot r} \epsilon_{ijk}
\frac{1}{2}(\chi^\dagger_{\bf q})_j  \bm{\sigma} \epsilon (\chi^\dagger_{-\bf q})_k\right)\cdot
\left( \sum_{\bf p} e^{i \bf p \cdot r} 
\frac{1}{2} \epsilon_{ilm}(\chi_{\bf -p})_l \epsilon \bm{\sigma}(\chi_{\bf p})_m\right) \nn \\
&-&\frac{1}{2} \int d^3 {\bf r} \, V^{(\bar 6)}(r) \left( \sum_{\bf q} e^{-i \bf q \cdot r} 
\frac{1}{\sqrt{2}} d^{(mn)}_{ij} (\chi^\dagger_{\bf q})_i   \epsilon (\chi^\dagger_{\bf -q})_j\right) 
\left( \sum_{\bf p} e^{i \bf p \cdot r} 
\frac{1}{\sqrt{2}} d^{(mn)}_{kl} (\chi_{\bf -p})_k \epsilon^T (\chi_{\bf p})_l\right) \nn \, ,
\eea 
where 
\bea
V^{(3)}(r) = -\frac{2}{3} \frac{\alpha_s}{r}, \qquad V^{(\bar 6)}(r) = \frac{1}{3} \frac{\alpha_s}{r} \, .
\eea
In Eq.~(\ref{s1}), we have left color indices explicit but suppressed spin indices. Note that the Fierz in
Eq.~(\ref{Fierz}) introduces four terms but two vanish due to Fermi statistics. Diquarks in the ${\bf  3} \,({\bf \bar 6})$
representation must be in a spin-1 (spin-0) state. Now we want to introduce composite fields which annihilate and 
create heavy diquarks. We define these fields as 
\bea \label{diq}
{\bf T}^i_{{\bf r}} &=& 
\sum_{\bf p} e^{i \bf p \cdot r} \frac{1}{2}\, \epsilon^{ijk} (\chi_{\bf -p})_j \epsilon \, \bm{\sigma} (\chi_{\bf p})_k \nn \\
\Sigma^{(mn)}_{\bf r} &=& \sum_{\bf p} e^{i \bf p \cdot r} \frac{1}{\sqrt{2}} \, 
d^{(mn)}_{ij} (\chi_{\bf-p})_i \epsilon^T (\chi_{\bf p})_j \nn \, .
\eea
The interaction terms can be made reparametrization invariant using a replacement analogous to that in Eq.~(\ref{rpi}).
Note that ${\bf T}^i_{\bf r}$ is a vector and $\Sigma^{(mn)}_{\bf r}$ is a scalar, as required by Fermi statistics. 
In the $Q\bar Q$, sector, we introduced  fields $S_{\bf r}$ and $O^a_{\bf r}$ that are bi-spinors which could be 
further reduced into their spin-0 and spin-1 components. If the diquarks are composed of two different heavy antiquarks,
Fermi statistics no longer correlates the color and spin quantum numbers. We would have to either use fields
in reducible representations of heavy quark spin $SU(2)$, as is usually done in pNRQCD
\cite{Brambilla:1999xf,Brambilla:2005yk}, or else introduce additional fields for spin-0 $\bf 3$ 
and spin-1 $\bf \bar 6$ diquarks. The diquark fields enter the theory via the Hubbard-Stratonovich trick:
\bea\label{DelL}
\Delta {\cal L} &=& 
\frac{1}{2} \int d^3 {\bf r} \, V^{(3)}(r) \left({\bf T}^{i\dagger}_{\bf r} -  \sum_{\bf q} e^{-i \bf q\cdot r} \epsilon_{ijk}
\frac{1}{2}(\chi^\dagger_{\bf q})_j  \bm{\sigma} \epsilon (\chi^\dagger_{\bf -q})_k  \right) \nn \\
&&\qquad \times
\left({{\bf T}}^i_{\bf r}-\sum_{\bf p} e^{i \bf p\cdot r} 
\frac{1}{2} \epsilon_{ilm}(\chi_{\bf -p})_l \epsilon \bm{\sigma}(\chi_{\bf p})_m\right)  \nn \\
&& + \frac{1}{2} \int d^3 {\bf r} \, V^{(\bar 6)}(r) \left(\Sigma^{(mn)\dagger}_{\bf r} - \sum_{\bf q} e^{-i \bf q \cdot r} \frac{1}{\sqrt{2}} \,
d^{(mn)}_{ij} (\chi_{\bf q})_i \epsilon (\chi_{\bf -q})_j \right) \nn \\
&&\qquad \times \left( \Sigma^{(mn)}_{\bf r}- \sum_{\bf p} e^{i \bf p\cdot r} \frac{1}{\sqrt{2}} \,
d^{(mn)}_{ij} (\chi_{\bf-p})_i \epsilon^T (\chi_{\bf p})_j  \right) \, .
\eea
The Feynman rules coupling diquarks to two heavy antiquarks from the interaction terms in Eq.~(\ref{DelL})
are given in Fig.~\ref{Fig5}. The Greek letters $\alpha$ and $\beta$ in these diagrams
are spin indices of the antiquarks. The Feynman rules for the $O(v)$ couplings of diquarks, two antiquarks and a gluon
which come from the reparametrization invariant extension of Eq.~(\ref{DelL}) are given in Fig.~\ref{Fig6}.

Diagrams analagous to those in Fig.~\ref{efv} give rise to electric dipole couplings of the diquarks but in our case  these couplings vanish.
Electric dipole couplings do not flip the spin of the heavy antiquark and therefore cannot couple diquarks of different spin such as 
${\bf T}^i_{\bf r}$ and $\Sigma^{(mn)}_{\bf r}$. If diquarks are composed of two different flavors of antiquarks, then  electric dipole
transitions between states with different color but identical spin are possible~\cite{Brambilla:2005yk}. Electric dipole moments of the 
${\bf T}^i_{\bf r}$ and $\Sigma^{(mn)}_{\bf r}$ vanish when the two antiquarks in the diquark have equal mass~\cite{Brambilla:2005yk}. 
Ref.~\cite{Brambilla:2005yk} has computed the electric dipole couplings in the case of non-identical antiquarks and we have checked that our 
methods reproduce their electric dipole couplings in this case.  

The chromomagnetic coupling of the diquarks is obtained by graphs like b) and c) of Fig.~\ref{efv}, with the  ${\bf p}\cdot{\bf A}$
coupling of the antiquarks replaced with the $\bm{\sigma}\cdot{\bf B}$ coupling in Eq.~(\ref{s1}). These diagrams are $O(v^2)$ and give rise
to the lowest order spin symmetry violating couplings of the diquarks. Other $O(v^2)$ diquark couplings that do not violate heavy quark 
spin symmetry can be found in Ref.~\cite{Brambilla:2005yk}. For the lowest order chromomagnetic coupling we find 
\bea
{\cal L}_{SSB} &=& -\frac{g}{2 m_Q}\int d^3 {\bf r} \,  i \, {{\bf T}}^{i\,\dagger}_{\bf r} \cdot{\bf B}^c \, T^c_{ij} \times{\bf T}^j_{\bf r}
- \frac{g}{\sqrt{2} m_Q}\int d^3 {\bf r} \,  \Sigma^{(mn)\dagger}_{\bf r} \, \epsilon_{ijk} \, \bar{T}^c_{lj} \, d^{(mn)}_{kl}{\bf B}^c \cdot 
{\bf T}^i_{\bf r} \, . 
\eea
More chromomagnetic couplings are possible if one considers diquarks composed of  two quark flavors. For instance, the $\Sigma^{(mn)}_{\bf r}$
field does not have a chromomagentic moment because it has spin 0, but if there are two flavors of heavy quark then the spin-1 
field in the $\bf \bar 6$ of color has a nonvanishing magnetic moment~\cite{Brambilla:2005yk}. 

This Lagrangian is not yet in the form of Eq.~(\ref{hqlag}). To get an effective action for local diquark fields of the form 
of Eq.~(\ref{hqlag}) we can formally expand the composite fields in the eigenfunctions of the leading order Hamiltonian.
For example, let
\bea
{{\bf T}}^i_{\bf r} =\sum_n{{\bf T}}^i_n \phi_n({\bf r}) \, ,
\eea 
where the $\phi_n({\bf r})$ are an orthonormal basis of energy eigenfunctions satisfying 
\bea\label{ev}
\left( -\frac{\bm{\nabla}_{\bf r}^2}{m_Q} + V^{(3)}(r)\right) \phi_n({\bf r}) = -\delta_n \phi_n({\bf r}) \, .
\eea
Then the Lagrangian for nonlocal composite field ${{\bf T}}^i_{\bf r}$ can be rewritten as a local 
Lagrangian with an infinite number of fields:
\bea \label{TL}
{\cal L}_{{\bf T}} &=& \int d^3 {\bf r} \, {{\bf T}}^\dagger_{\bf r}\left(i D_0 
+ \frac{\bm{\nabla}_{\bf r}^2}{m_Q} -V^{(3)}(r)\right){{\bf T}}_{\bf r} 
-\frac{g}{2 m_Q}\int d^3 {\bf r} \,  i \, {{\bf T}}^{\dagger}_{\bf r} \cdot{\bf B}  \times{\bf T}_{\bf r}
\nn \\
&=& \sum_n {{\bf T}}^\dagger_{n} (i D_0 + \delta_n){{\bf T}}_{n} - \frac{g}{2 m_Q} i \,\sum_n \, {{\bf T}}^{\dagger}_{n} \cdot{\bf B}  
\times{\bf T}_{n} \, .
\eea 
(Here we have suppressed $SU(3)$ indices.)
This assumes that the spectrum of the two-body Hamiltonian in Eq.~(\ref{ev}) is discrete. This is not the case within perturbation theory 
but must be true nonperturbatively in a confining theory like QCD. There are many heavy diquark fields and each has a different residual momentum
given by the binding energy of the diquark in the nonrelativistic potential in Eq.~(\ref{ev}). If one is interested in the properties of the 
ground state doubly heavy baryon only the lowest lying diquark field is relevant. Diquark fields corresponding to excited states of the 
diquark have an excitation energy of $O(m_Q v^2)$ as do the low lying states of the $\Sigma^{(mn)}_{\bf r}$ fields, whose constituents 
feel a repulsive potential at short distances and therefore must be heavier by $O(m_Q v^2)$ than the ground state of the diquark in the $\bf 3$ 
representation. For $m_Q v^2 \gg \Lambda_{\rm QCD}$ one can integrate out these degrees of freedom and one is left with a diquark Lagrangian 
of the form of Eq.~(\ref{hqlag}) with an additional residual mass term due to the diquark binding energy. This 
residual mass term does not affect the leading hyperfine splitting prediction of Eq.~(\ref{hf}) at $O(1/m_Q)$.

\section{Soft Gluons}

In this section we discuss the role of soft gluons in the effective theory for composite fields. In vNRQCD 
soft gluons appear at leading order in $v$, where the coupling to the quarks is given by
\bea\label{soft}
{\cal L}_{soft} = -\frac{g^2}{2} \sum_{{\bf p,p^\prime},q,q^\prime}  
\psi_{p^\prime}^\dagger [A_q^\mu, A_{q^\prime}^\nu] U_{\mu \nu}({\bf p,p^\prime},q,q^\prime) \psi_p + (\psi \to \chi) \, .
\eea
Here $U_{\mu \nu}({\bf p,p^\prime},q,q^\prime)$ is a complicated function of the momenta which  can be found in
Refs.~\cite{Luke:1999kz,Manohar:2000hj}. At leading order in $v$ the soft coupling respects heavy quark spin symmetry, with symmetry breaking
contributions to $U({\bf p,p^\prime},q,q^\prime)$ suppressed by one power of $v$. There are also soft couplings with an anti-commutator
structure, $\{A_q^\mu, A_{q^\prime}^\nu \}$, whose leading contributions are suppressed by one power of $v$ and preserve heavy quark spin symmetry. 

Integrating out the heavy quarks via the Hubbard-Stratonovich transformation results in a coupling of the composite fields to soft gluons.
However, soft gluons carry energy and momentum of $O(m_Q v)$ so emission of soft gluons couples  diquark fields with different momentum
labels. Because the effective theory is applied to diquarks which are bound in a baryon strong interactions can only change the composite
field's total momentum by  $O(m_Q v^2)$ or smaller. Thus soft gluons cannot appear as external degrees of freedom. They do, however, appear
in loops where they can form a tadpole,  give a higher order (in  $\alpha_s(m_Q v)$) correction to the
potential, or  connect to an external usoft spectator. This last contribution is important for doubly heavy baryons since there is a valence
usoft quark, and couplings of the diquark to the usoft quark via soft gluons could in principle lead to spin symmetry breaking interactions.  

\begin{figure}[!t]
  \centerline{\epsfysize=4.0truecm \epsfbox[225 560 380 670]{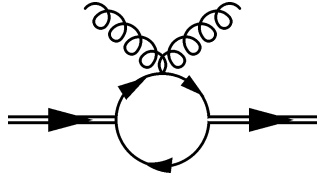}  }
 {\tighten
\caption[1]{Coupling a composite field to soft gluons.}
\label{Fig7} }
\end{figure}

One can try to derive a coupling of soft gluons to composite fields by evaluating a diagram like that in  Fig.~\ref{Fig7}, where the gluons
are soft rather than usoft. However, such a diagram is problematic because the soft gluon coupling depends on the label momenta, and this
coupling becomes singular in the momentum region of interest. To be specific consider Fig.~\ref{Fig7}. Let one gluon carry incoming label
three-momentum ${\bf q}$ and the other outgoing three-momentum ${\bf q^\prime}$, and let the heavy quark and anti-quark coupling to the
initial composite field have three momentum ${\bf l}$ and $-{\bf l}$, respectively. The heavy quark coupling to the final state composite
field has label momentum ${\bf l +q -q^\prime}$. Since the final heavy quark and antiquark have total nonvanishing label momentum $\bf q
-q^\prime$  the coupling to the final state composite fields  vanishes except at ${\bf q -q^\prime} = 0$. When ${\bf q -q^\prime} = 0$ the
soft coupling is singular since $U_{\mu \nu}({\bf p,p^\prime},q,q^\prime)$ has terms proportional to  $(q^0,{\bf q}^i)/|{\bf p-p^\prime}|^2$,
and $|{\bf q - q^\prime}|=|{\bf p - p^\prime}|$  by label momentum conservation.

The singular coupling arises because one is integrating out a gluon exchanged in the t-channel between the
heavy quark and soft gluon fields appearing in Eq.~(\ref{soft}). For generic $\bf p$ and $\bf p^\prime$ 
this gluon has energy $O(m_Q v^2)$ and momentum $O(m_Q v)$. This mode is off-shell and should be integrated out of the theory.
But when $\bf p=\bf p^\prime$, this mode has energy and momentum $O(m_q v^2)$ and is therefore a usoft mode 
which should not be integrated out of the theory. Therefore, the singular interaction can be removed from the 
theory by adopting the prescription that $\bf p = \bf p^\prime$ (and $q=q^\prime$) is excluded from the sum
in Eq.~(\ref{soft}). Then there will be no coupling of soft gluons to diquarks.

Suppose as an alternative we adopt an ad-hoc prescription for dealing with the singularity, such as  replacing $(\bf p - \bf p^\prime)^2$
with $({\bf p} - {\bf p^\prime})^2 + \Lambda^2$, where $\Lambda^2$  is an IR regulator, in the denominator of singular terms in $U_{\mu
\nu}({\bf p,p^\prime},q,q^\prime)$. Then there will be a  coupling of diquarks to soft gluons. However, the loop graphs with this  coupling
will vanish because the integrals over soft momentum are scaleless, since there is no external soft momentum, and scaleless integrals are
set to zero in dimensional regularization.

Integrating out the large scale $m_Q$ with external heavy and usoft 
quarks give rises to four quark operators of the form~\cite{Bauer:1997gs}
\bea
{\cal L} &=&\frac{c_1}{8 m_Q^2} \psi^\dagger T^a \psi \, \bar q \vsl T^a q 
+ \frac{c_2}{8 m_Q^2}  \psi^\dagger  \sigma^i T^a \bm{\sigma} \psi \cdot \bar q \, \bm{\gamma} \gamma_5 T^a q \nn \\
&+& \frac{c_3}{8 m_Q^2}  \psi^\dagger  \psi \, \bar q \vsl q 
+ \frac{c_4}{8 m_Q^2}  \psi^\dagger  \bm{\sigma} \psi  \cdot \bar q \,  \bm{\gamma} \gamma_5 q \, .
\eea
The operators with coefficients $c_2$ and $c_4$ violate heavy quark spin symmetry and can induce 
heavy quark spin symmetry violating couplings of the diquarks in the effective theory. Using the methods developed 
in this paper it is is straightforward to derive these couplings. For example we obtain
\bea
{\cal L} &=&  \frac{c_2}{8 m_Q^2}\int d^3 {\bf r} \,  i \, {{\bf T}}^{i \, \dagger}_{\bf r} \cdot \bar q \,\bm{\gamma} \gamma_5 T^c q  \, \bar T^c_{ij} 
\times{\bf T}^j_{\bf r} \, .
\eea
Such operators are suppressed by $O(\alpha_s(m_Q)^2 v^2)$ relative to the leading hyperfine splitting term.
The matching  coefficients, $c_i$, start at $O(\alpha_s^2)$ and are calculated at the scale $m_Q$. The usoft quark
bilinear, $\bar q \,\bm{\gamma}\gamma_5 q$, is $O(v^6)$ while usoft chromomagnetic field, $\bf B$, is $O(v^4)$ in vNRQCD power counting.
Finally, there are corrections to the diquark chromomagnetic coupling coming from vNRQCD operators
coupling four heavy quark fields and a usoft chromomagnetic field. The coefficients of these operators
are suppressed by two powers of $\alpha_s$ and should give the leading correction to the prediction 
for the hyperfine splittings of doubly heavy baryons.

\section{Conclusions}

Baryons with two heavy antiquarks can be understood as a point-like diquark, composed of the two heavy antiquarks, interacting with
the light antiquark. The reason is that the heavy quarks form a nonrelativistic bound state of size $r \sim 1/(m_Q v)$, which is much smaller
than $1/\Lambda_{\rm QCD}$ the wavelength of the light degrees of freedom. As a consequence there is a heavy diquark-quark symmetry which
relates doubly heavy baryons to heavy mesons. However, the dynamics of the doubly heavy baryons is described by NRQCD not HQET. In this paper
we show how heavy diquark-quark symmetry arises within the framework of NRQCD. Starting from the $Q\bar{Q}$ sector of the vNRQCD Lagrangian
we use a Hubbard-Stratanovich transformation to derive a Lagrangian in terms of color-singlet and color-octet degrees of freedom. The
Lagrangian we derive corresponds to the leading pNRQCD action. This analysis clarifies the relationship between vNRQCD and pNRQCD. We extend
the formalism to the $QQ$ sector, where we recast vNRQCD into a theory of diquarks in the $\bf \bar{3}$ and $\bf 6$ representation of color.
We include in our Lagrangian the leading symmetry breaking terms, which  involve the chromomagnetic fields. In our derivation we  consider
the effects of soft gluons in vNRQCD, and argue that they only give corrections to potentials. In particular we claim that soft gluons do not
contribute to the coupling of spectator quarks to the diquarks since any possible contributions vanishes in dimensional regularization. As a
consequence operators coupling light quarks to diquarks only come from integrating out the scale $m_Q$, and are suppressed by $\alpha_s(m_Q)^2
v^2$ relative to the leading spin symmetry breaking term. 

To arrive at an effective action for diquark fields that is of the form of the HQET Lagrangian we expand the diquark field in eigenstates of
the Hamiltonian. When   $m_Q v^2 \gg \Lambda_{\rm QCD}$ we are justified in keeping only the lowest energy mode, and thereby obtain an HQET
like Lagrangian with an additional residual mass term due to the diquark binding energy. This additional term does not affect the leading
hyperfine splitting prediction. Thus we are able to derive spin symmetry relations between heavy mesons and double heavy baryons from NRQCD.
These spin symmetry relations are satisified at the  $30$\% level for the charm sector even though the required hierarchy of scales  $m_Q \gg
m_Q v \gg m_Q v^2 \gg \Lambda_{\rm QCD}$ is not well satisfied for the  $c c$ system. The deviations from the spin symmetry
relations are of the expected size. This makes a $v$ expansion for this observable reasonable,
and motivates a study of the higher order corrections which will be presented in a future publication.

\begin{acknowledgments}  

We thank I. Stewart, A. Hoang, N. Brambilla and A. Vairo for useful discussions.
We also thank the Institute for Nuclear Theory for their hospitality during the completion of this work.
T.M. is supported in part by Department of Energy grants DE-FG02-05ER41368, DE-FG02-05ER64101 and DE-AC05-84ER40150.
 \end{acknowledgments}


\end{document}